\documentclass[12pt,twoside,english,refpage,intoc,bibliography=totoc,index=totoc,BCOR7.5mm,captions=tableheading]{article}
\usepackage{lmodern}

\usepackage[T1]{fontenc}
\usepackage[latin9]{inputenc}
\usepackage{color}
\usepackage{babel}
\usepackage{dsfont}
\usepackage{amsmath}
\usepackage{geometry}
\usepackage[bookmarks=true,bookmarksnumbered=true,bookmarksopen=false,
 breaklinks=false,pdfborder={0 0 1},backref=false,colorlinks=true]
 {hyperref}
\hypersetup{pdftitle={The LyX User's Guide},
 pdfauthor={LyX Team},
 pdfsubject={LyX},
 pdfkeywords={LyX},
 linkcolor=black, citecolor=black, urlcolor=blue, filecolor=blue, pdfpagelayout=OneColumn, pdfnewwindow=true, pdfstartview=XYZ, plainpages=false}

\makeatletter
\numberwithin{equation}{section}

\usepackage{babel}
\usepackage[figure]{hypcap}

\usepackage{fancyhdr}

\@ifundefined{extrarowheight}{\usepackage{array}}{}
\usepackage{braket}
\makeatother

\begin{document}
\title{Sliver frame stubs in OSFT via auxiliary fields}
\author{Georg Stettinger}
\maketitle
\begin{center}
\textcolor{black}{{CEICO, Institute of Physics of the Czech Academy
of Sciences, }\\
\textcolor{black}{{} Na Slovance 2, 182 00 Prague 8, Czech Republic} }
\par\end{center}
\begin{abstract}
In this short paper we want to generalize some recent concepts related
to stubs in open string field theory. First, we modify the auxiliary
field method by Erbin and Firat \cite{Erbin2023} to non-BPZ even
sliver frame stubs. We then also show that the construction is consistent
at the full quantum level without any additional assumptions. Finally,
we apply the method explicitly to the tachyon vacuum and the simplest
identity-like solution.

\tableofcontents{}
\end{abstract}

\section{Introduction }

Deforming Witten's open string field theory with stubs has a couple
of interesting motivations and features. First of all, the stubbed
OSFT builds a bridge between the structures of open and closed string
field theory and will also be of relevance in a combined open-closed
SFT. Second, it can help us to understand general properties of classical
solutions to $A_{\infty}$-type theories. Moreover, it is also expected
to tame certain singularities present in Witten theory, for example
associated to identity-like solutions. 

The starting point is the following: Instead of Witten theory defined
by the action 
\begin{equation}
S\left(\Psi\right)=-\frac{1}{2}\omega\left(\Psi,Q\Psi\right)-\frac{1}{3}\omega\left(\Psi,m_{2}\left(\Psi,\Psi\right)\right)
\end{equation}
we consider a modification of the two-product
\begin{equation}
m_{2}\left(\cdot,\cdot\right)\rightarrow M_{2}\left(\cdot,\cdot\right)=e^{-\lambda L_{0}}m_{2}\left(e^{-\lambda L_{0}}\cdot,e^{-\lambda L_{0}}\cdot\right).
\end{equation}
This new product $M_{2}$ is not associative anymore, hence to ensure
gauge invariance, we need to introduce higher products $M_{n\geq3}$
to obtain an $A_{\infty}$-algebra as described in detail in \cite{Schnabl:2023dbv,Schnabl2024}.
Erbin and Firat have shown in \cite{Erbin2023} that it is also possible
to obtain the stubbed $A_{\infty}$-theory from a cubic theory containing
an auxiliary string field. By integrating out the auxiliary field
in a specific way one can either get the stubbed theory or go back
to Witten theory. Their results were then generalized to the full
quantum level by Maccaferri et al in \cite{Maccaferri2024}. 

In both works it is a necessary condition for the stub operator to
be BPZ-even, which does not include the sliver frame stub $e^{-\lambda\mathcal{L}_{0}}$.
Since most analytical solutions are formulated in the sliver frame,
it is of interest to use $e^{-\lambda\mathcal{L}_{0}}$ since many
algebraic manipulations become much simpler. While the higher products
and the maps relating solutions were already constructed in \cite{Schnabl2024},
in this work we want to focus on the auxiliary field method and the
quantum theory. We point out that geometric aspects of sliver frame
stubs are actually very subtle and some aspects are adressed in \cite{Schnabl2024},
see also \cite{Kiermaier:2007jg}. In this paper we focus solely on
algebraic aspects. 

The paper is organized as follows: Section 2.1 is just a summary of
the work by Erbin and Firat which is included for self-consistency
and does not contain any new results. In section 2.2 we work out the
generalization to the sliver frame and also discuss an alternative
way to obtain Witten theory by integrating out the auxiliary field.
Section 2.3 is then devoted to cohomomorphisms and how to obtain classical
solutions of the stubbed equations of motion. In section 3 we show
that our construction also works at the full quantum level without
any additional constraints. Finally, in section 4 we apply the formalism
to two known analytic solutions, namely the tachyon vacuum and one
identity-like solution. 

\section{Sliver frame stubs as an auxiliary field }

\subsection{Ordinary stubs as an auxiliary field }

In this section we want to give an overview of the results by Erbin
and Firat \cite{Erbin2023}, who managed to decribe the stubbed $A_{\infty}$-theory
in a cubic way using an additional auxiliary string field. By integrating
out the auxiliary field via homotopy transfer one obtains the infinitely
many higher products. 

The first step is to double the Hilbert space and turn the string
field into a two-vector:

\begin{equation}
\Psi\in\mathcal{H}\,\,\,\,\,\,\rightarrow\,\,\,\,\,\,\Phi=\begin{pmatrix}\Psi\\
\Sigma
\end{pmatrix}\in\mathcal{H}\oplus\mathcal{H}.
\end{equation}
The proposed action for the combined string field $\Phi$ reads explicitly
\begin{align}
S= & \,-\frac{1}{2}\omega\left(\Psi,Q\Psi\right)-\frac{1}{2}\omega\left(\Sigma,\frac{Q}{1-e^{-2\lambda L_{0}}}\Sigma\right)\nonumber \\
 & \,+\frac{1}{3}\omega\left(\Sigma-e^{-\lambda L_{0}}\Psi,m_{2}\left(\Sigma-e^{-\lambda L_{0}}\Psi,\Sigma-e^{-\lambda L_{0}}\Psi\right)\right)\label{eq:atakan action}
\end{align}
and is invariant under the infinitesimal gauge transformations 
\begin{equation}
\delta\Psi=Q\Lambda_{1}+e^{-\lambda L_{0}}m_{2}\left(\Sigma-e^{-\lambda L_{0}}\Psi,\Lambda_{2}-e^{-\lambda L_{0}}\Lambda_{1}\right)+e^{-\lambda L_{0}}m_{2}\left(\Lambda_{2}-e^{-\lambda L_{0}}\Lambda_{1},\Sigma-e^{-\lambda L_{0}}\Psi\right),
\end{equation}
\begin{equation}
\delta\Sigma=Q\Lambda_{2}+\left(e^{-2\lambda L_{0}}-1\right)\left(m_{2}\left(\Sigma-e^{-\lambda L_{0}}\Psi,\Lambda_{2}-e^{-\lambda L_{0}}\Lambda_{1}\right)+m_{2}\left(\Lambda_{2}-e^{-\lambda L_{0}}\Lambda_{1},\Sigma-e^{-\lambda L_{0}}\Psi\right)\right).
\end{equation}
Here, $\Lambda_{1}$ and $\Lambda_{2}$ are two independent gauge
parameters both of ghost number zero. The algebraic ingredients giving
rise to this action are a differential 
\begin{equation}
V_{1}\left(\Phi\right)=\begin{pmatrix}Q\Psi\\
Q\Sigma
\end{pmatrix},
\end{equation}
an associative product 
\begin{equation}
V_{2}\left(\Phi_{1},\Phi_{2}\right)=\begin{pmatrix}e^{-\lambda L_{0}}m_{2}\left(\Sigma_{1}-e^{-\lambda L_{0}}\Psi_{1},\Sigma_{2}-e^{-\lambda L_{0}}\Psi_{2}\right)\\
\left(e^{-2\lambda L_{0}}-1\right)m_{2}\left(\Sigma_{1}-e^{-\lambda L_{0}}\Psi_{1},\Sigma_{2}-e^{-\lambda L_{0}}\Psi_{2}\right)
\end{pmatrix}
\end{equation}
and a bilinear form 
\begin{equation}
\Omega\left(\Phi_{1},\Phi_{2}\right)=\omega\left(\Psi_{1},\Psi_{2}\right)+\omega\left(\Sigma_{1},\frac{1}{1-e^{-2\lambda L_{0}}}\Sigma_{2}\right).
\end{equation}
Together, $V_{1}$, $V_{2}$ and $\Omega$ form a cyclic differential
graded algebra. We see that $\Omega$ is ill-defined on states in
the kernel of $L_{0}$. To solve that, we just assume that $\Sigma$
can only take values outside of $\text{ker}\,L_{0}$, such that our
Hilbert space is effectively $\mathcal{H}\oplus\left(\mathcal{H}\setminus\text{ker }L_{0}\right)$.
In \cite{Maccaferri2024} it has been shown that this action can be
obtained by adding a trivial, non-propagating field to Witten theory
and then performing an orthogonal rotation amongst the two fields. 

To obtain the non-polynomial stubbed action one has to integrate out
the auxiliary field $\Sigma$ using homotopy transfer. We therefore
define the natural inclusion and projection maps 
\begin{equation}
i_{S}=\begin{pmatrix}1\\
0
\end{pmatrix},\,\,\,\,\,\,\,\,p_{S}=\begin{pmatrix}1 & 0\end{pmatrix}\label{eq:natural i p}
\end{equation}
and choose for the homotopy 
\begin{equation}
h_{S}=\begin{pmatrix}0 & 0\\
0 & -\frac{b_{0}}{L_{0}}
\end{pmatrix}
\end{equation}
which fulfills the Hodge-Kodaira relation 
\begin{equation}
h_{S}V_{1}+V_{1}h_{S}=i_{S}p_{S}-1.
\end{equation}
Those maps give rise to a special deformation retract which means
they obey $p_{S}i_{S}=1$ as well as the side conditions 
\begin{equation}
h_{S}i_{S}=0,\,\,\,\,\,\,\,\,\,p_{S}h_{S}=0,\,\,\,\,\,\,\,\,\,\,h_{S}h_{S}=0.
\end{equation}
In \cite{Schnabl:2023dbv} it has been shown that those conditions
are actually not necessary to perform homotopy transfer, only the
Hodge-Kodaira relation is needed. However, if they are obeyed, we
can follow the standard procedure known in the literature without
any modifications. Applying the tensor trick (see for instance \cite{Vosmera:2020jyw})
and adding $\mathbf{V_{2}}$ as a perturbation, we get from the homological
perturbation lemma the new products as 
\begin{equation}
M_{n}=\pi_{1}\mathbf{p_{S}}\mathbf{V_{2}}\left(1-\mathbf{h_{S}V_{2}}\right)^{-1}\mathbf{i_{S}}\pi_{n},\,\,\,\,\,\,\,\,\,\,\,\,\,\,\,n\geq2.\label{eq:higher products}
\end{equation}
or more explicitly
\begin{align}
M_{2}\left(\cdot,\cdot\right) & =e^{-\lambda L_{0}}m_{2}\left(e^{-\lambda L_{0}}\cdot,e^{-\lambda L_{0}}\cdot\right)\nonumber \\
M_{3}\left(\cdot,\cdot,\cdot\right) & =e^{-\lambda L_{0}}m_{2}\left(e^{-\lambda L_{0}}\cdot,\frac{e^{-2\lambda L_{0}}-1}{L_{0}}b_{0}m_{2}\left(e^{-\lambda L_{0}}\cdot,e^{-\lambda L_{0}}\cdot\right)\right)\nonumber \\
 & \,+\,e^{-\lambda L_{0}}m_{2}\left(\frac{e^{-2\lambda L_{0}}-1}{L_{0}}b_{0}m_{2}\left(e^{-\lambda L_{0}}\cdot,e^{-\lambda L_{0}}\cdot\right),e^{-\lambda L_{0}}\cdot\right)\\
 & \vdots\nonumber 
\end{align}
In \cite{Erbin2023} it has been shown that they are exactly equivalent
to the higher products constructed in \cite{Schnabl:2023dbv}. 

If we want to get back Witten theory, we have to project on the subspace
of $\mathcal{H}\oplus\left(\mathcal{H}\setminus\text{ker }L_{0}\right)$
given by elements of the form $\begin{pmatrix}\Psi\\
-2\sinh\left(\lambda L_{0}\right)\Psi
\end{pmatrix}.$ The inclusion, projection and homotopy defined in \cite{Erbin2023}
read 
\begin{align}
 & i_{W}=\begin{pmatrix}1\\
-2\sinh\left(\lambda L_{0}\right)
\end{pmatrix},\,\,\,\,\,\,\,\,\,p_{W}=\frac{1}{1+4\sinh^{2}\left(\lambda L_{0}\right)}\begin{pmatrix}1 & -2\sinh\left(\lambda L_{0}\right)\\
-2\sinh\left(\lambda L_{0}\right) & 4\sinh^{2}\left(\lambda L_{0}\right)
\end{pmatrix},\label{eq:i' and p'}
\end{align}
\begin{equation}
h_{W}=-\frac{b_{0}}{L_{0}}\frac{1}{1+4\sinh^{2}\left(\lambda L_{0}\right)}\begin{pmatrix}4\sinh^{2}\left(\lambda L_{0}\right) & 2\sinh\left(\lambda L_{0}\right)\\
2\sinh\left(\lambda L_{0}\right) & 1
\end{pmatrix}.
\end{equation}
Again, they form a special deformation retract so homotopy transfer
works straightforwardly. Especially, the ``magical relation'' 
\begin{equation}
h_{W}V_{2}=0\label{eq:magical relation}
\end{equation}
holds which can be seen by direct calculation. In fact, by writing
out (\ref{eq:higher products}) in components and using the side conditions,
one can see that $h_{W}V_{2}$ appears in all terms contained in the
products with more then two inputs. This means that all the higher
products vanish identically and we are left with only a two-product
and, as a result, a cubic theory. If we write down the action with
input fields $i_{W}\Psi$ and the perturbed product we get 
\begin{align}
S & =-\frac{1}{2}\Omega\left(i_{W}\Psi,V_{1}\left(i_{W}\Psi\right)\right)-\frac{1}{3}\Omega\left(i_{W}\Psi,p_{W}V_{2}\left(i_{W}\Psi,i_{W}\Psi\right)\right)\nonumber \\
 & =-\frac{1}{2}\omega\left(e^{\lambda L_{0}}\Psi,Qe^{\lambda L_{0}}\Psi\right)-\frac{1}{3}\omega\left(e^{\lambda L_{0}}\Psi,m_{2}\left(e^{\lambda L_{0}}\Psi,e^{\lambda L_{0}}\Psi\right)\right)
\end{align}
which is exactly the Witten action up to the field redefinition $\Psi\rightarrow e^{\lambda L_{0}}\Psi$.

We want to point out that the Witten product $m_{2}$ was not directly
obtained from homotopy transfer: First, the $p_{W}$ in (\ref{eq:i' and p'})
is a two-by-two matrix, hence the target space of the homotopy transfer
is still $\mathcal{H}\oplus\left(\mathcal{H}\setminus\text{ker }L_{0}\right)$.
Only after writing down the action using $\Omega$ and undoing the
field redefinition $\Psi\rightarrow e^{\lambda L_{0}}\Psi$ it is
possible to read off $m_{2}$. In the next section we will give a
more direct way of obtaining Witten theory. 

\subsection{Generalization to sliver frame stubs}

Now we turn to the actual task and replace the stub operator by its
sliver frame counterpart. The whole construction above was only valid
for BPZ-even stub operators, hence if we want to use $e^{-\lambda\mathcal{L}_{0}}$
we need to generalize it. We propose to use the following product
and bilinear form (the differential will be unchanged)\footnote{From now on, primed objects always refer to the sliver frame stubs.}:
\begin{equation}
V_{2}'\left(\Phi_{1},\Phi_{2}\right)=\begin{pmatrix}e^{-\lambda\mathcal{L}_{0}^{*}}m_{2}\left(\Sigma_{1}-e^{-\lambda\mathcal{L}_{0}}\Psi_{1},\Sigma_{2}-e^{-\lambda\mathcal{L}_{0}}\Psi_{2}\right)\\
\left(e^{-\lambda\mathcal{L}_{0}}e^{-\lambda\mathcal{L}_{0}^{*}}-1\right)m_{2}\left(\Sigma_{1}-e^{-\lambda\mathcal{L}_{0}}\Psi_{1},\Sigma_{2}-e^{-\lambda\mathcal{L}_{0}}\Psi_{2}\right)
\end{pmatrix}
\end{equation}
\begin{equation}
\Omega'\left(\Phi_{1},\Phi_{2}\right)=\omega\left(\Psi_{1},\Psi_{2}\right)+\omega\left(\Sigma_{1},\frac{1}{1-e^{-\lambda\mathcal{L}_{0}}e^{-\lambda\mathcal{L}_{0}^{*}}}\Sigma_{2}\right).\label{eq:new symplectic form}
\end{equation}
One can show straightforwardly that the new product is again associative
and cyclic with respect to $\Omega'$, so all algebraic relations
are preserved. The new action reads 
\begin{align}
S'= & \,-\frac{1}{2}\omega\left(\Psi,Q\Psi\right)-\frac{1}{2}\omega\left(\Sigma,\frac{Q}{1-e^{-\lambda\mathcal{L}_{0}}e^{-\lambda\mathcal{L}_{0}^{*}}}\Sigma\right)\label{eq:curly auxiliary action}\\
 & \,+\frac{1}{3}\omega\left(\Sigma-e^{-\lambda\mathcal{L}_{0}}\Psi,m_{2}\left(\Sigma-e^{-\lambda\mathcal{L}_{0}}\Psi,\Sigma-e^{-\lambda\mathcal{L}_{0}}\Psi\right)\right)\nonumber 
\end{align}
and is invariant under the gauge transformations 
\begin{equation}
\delta\Psi=Q\Lambda_{1}+e^{-\lambda\mathcal{L}_{0}^{*}}\left(m_{2}\left(\Sigma-e^{-\lambda\mathcal{L}_{0}}\Psi,\Lambda_{2}-e^{-\lambda\mathcal{L}_{0}}\Lambda_{1}\right)+m_{2}\left(\Lambda_{2}-e^{-\lambda\mathcal{L}_{0}}\Lambda_{1},\Sigma-e^{-\lambda\mathcal{L}_{0}}\Psi\right)\right),
\end{equation}
\begin{equation}
\delta\Sigma=Q\Lambda_{2}+\left(e^{-\lambda\mathcal{L}_{0}}e^{-\lambda\mathcal{L}_{0}^{*}}-1\right)\left(m_{2}\left(\Sigma-e^{-\lambda\mathcal{L}_{0}}\Psi,\Lambda_{2}-e^{-\lambda\mathcal{L}_{0}}\Lambda_{1}\right)+m_{2}\left(\Lambda_{2}-e^{-\lambda\mathcal{L}_{0}}\Lambda_{1},\Sigma-e^{-\lambda\mathcal{L}_{0}}\Psi\right)\right).
\end{equation}
The operator $O=:\frac{1}{1-e^{-\lambda\mathcal{L}_{0}}e^{-\lambda\mathcal{L}_{0}^{*}}}$
appearing in the kinetic term actually deserves attention. First of
all, we can use the formulas of \cite{Schnabl2024,Schnabl:2005gv}
to write 
\begin{equation}
e^{-\lambda\mathcal{L}_{0}}e^{-\lambda\mathcal{L}_{0}^{*}}=e^{\left(e^{-\lambda}-1\right)\left(\mathcal{\mathcal{L}}_{0}+\mathcal{L}_{0}^{*}\right)}\label{L_hat exponential 1}
\end{equation}
so in fact $O$ is a function of $\hat{\mathcal{L}}=\mathcal{\mathcal{L}}_{0}+\mathcal{L}_{0}^{*}$.
In general, the exponential of $\hat{\mathcal{L}}$ acts on an arbitrary
string field as 
\begin{equation}
e^{-t\hat{\mathcal{\mathcal{L}}}}\Psi=e^{-tK}\Psi e^{-tK},\label{eq:L_hat exponential 2}
\end{equation}
hence the action of $O$ can be defined via its Laplace transform.
One state in the kernel of $\hat{\mathcal{L}}$ for which applying
$O$ could be problematic is the sliver state $e^{-\infty K}$. However,
this state is of formal nature and singular in many senses so we just
choose to exclude it from our Hilbert space. We will see that in the
end, after integrating out the auxiliary field, $O$ does not appear
anymore. 

The homotopy transfer to the stubbed theory is essentially unchanged
and again generated by (\ref{eq:natural i p}). For the homotopy,
the most compatible choice is 
\begin{equation}
h'_{S}=\begin{pmatrix}0 & 0\\
0 & -\frac{\mathcal{B}_{0}+\mathcal{B}_{0}^{*}}{\mathcal{L}_{0}+\mathcal{L}_{0}^{*}}
\end{pmatrix}:\label{eq:h in B-hat}
\end{equation}
The reason is that in the higher products $h'_{S}$ always acts on
the output of $V'_{2}$ such that the expression 
\begin{equation}
h'_{S}\left(e^{-\lambda\mathcal{L}_{0}}e^{-\lambda\mathcal{L}_{0}^{*}}-1\right)=h'_{S}\left(e^{\left(e^{-\lambda}-1\right)\left(\mathcal{\mathcal{L}}_{0}+\mathcal{L}_{0}^{*}\right)}-1\right)
\end{equation}
appears. Hence, with this choice, $h'_{S}V'_{2}$ is manifestly non-singular
and well-defined on all states. The new higher products one receives
are then explicitly given by 
\begin{align}
M'_{2}\left(\cdot,\cdot\right) & =e^{-\lambda\mathcal{\mathcal{L}}_{0}^{*}}m_{2}\left(e^{-\lambda\mathcal{\mathcal{L}}_{0}}\cdot,e^{-\lambda\mathcal{\mathcal{L}}_{0}}\cdot\right),\nonumber \\
M'_{3}\left(\cdot,\cdot,\cdot\right) & =e^{-\lambda\mathcal{\mathcal{L}}_{0}^{*}}m_{2}\left(e^{-\lambda\mathcal{\mathcal{L}}_{0}}\cdot,\frac{e^{\left(e^{-\lambda}-1\right)\left(\mathcal{\mathcal{L}}_{0}+\mathcal{L}_{0}^{*}\right)}-1}{\left(\mathcal{\mathcal{L}}_{0}+\mathcal{L}_{0}^{*}\right)}\left(\mathcal{\mathcal{B}}_{0}+\mathcal{B}_{0}^{*}\right)m_{2}\left(e^{-\lambda\mathcal{\mathcal{L}}_{0}}\cdot,e^{-\lambda\mathcal{\mathcal{L}}_{0}}\cdot\right)\right)\,\nonumber \\
 & \,\,\,\,\,\,+e^{-\lambda\mathcal{\mathcal{L}}_{0}^{*}}m_{2}\left(\frac{e^{\left(e^{-\lambda}-1\right)\left(\mathcal{\mathcal{L}}_{0}+\mathcal{L}_{0}^{*}\right)}-1}{\left(\mathcal{\mathcal{L}}_{0}+\mathcal{L}_{0}^{*}\right)}\left(\mathcal{\mathcal{B}}_{0}+\mathcal{B}_{0}^{*}\right)m_{2}\left(e^{-\lambda\mathcal{\mathcal{L}}_{0}}\cdot,e^{-\lambda\mathcal{\mathcal{L}}_{0}}\cdot\right),e^{-\lambda\mathcal{\mathcal{L}}_{0}}\cdot\right).\\
 & \vdots\nonumber 
\end{align}
They are identical to the products with curly stubs derived in \cite{Schnabl2024}.
We also note that the operator $O$ does not appear anymore and all
quantities are well-defined.

The transfer to Witten theory is actually a bit more subtle: It is
possible to construct maps $\tilde{i}_{W}$ and $\tilde{p}_{W}$ which
lead directly to Witten theory, without any field redefinition being
necessary. Let us first see how this works for the ordinary stubs
where we define 
\begin{equation}
\tilde{i}_{W}=\begin{pmatrix}e^{-\lambda L_{0}}\\
e^{-2\lambda L_{0}}-1
\end{pmatrix},\,\,\,\,\,\,\,\,\,\,\tilde{p}{}_{W}=\begin{pmatrix}e^{-\lambda L_{0}} & -1\end{pmatrix}.\label{eq:cool i and p}
\end{equation}
They do not satisfy the typical properties of an inclusion and a projection,
but this is by no means necessary; they only need to commute with
the differential and obey the Hodge-Kodaira relation, which can be
solved by 
\begin{equation}
\tilde{h}{}_{W}=\frac{b_{0}}{L_{0}}\begin{pmatrix}e^{-2\lambda L_{0}}-1 & -e^{-\lambda L_{0}}\\
e^{-3\lambda L_{0}}-e^{-\lambda L_{0}} & -e^{-2\lambda L_{0}}
\end{pmatrix}.
\end{equation}
Now the maps $\tilde{i}_{W}$, $\tilde{p}_{W}$ and $\tilde{h}_{W}$
even satisfy all the SDR-conditions. The tensorial structure of $\tilde{i}$
and $\tilde{p}$ implies that the resulting theory is defined on just
one copy of $\mathcal{H}$, so we can directly calculate all the products
without computing the action first. Most importantly, the magical
relation $\tilde{h}{}_{W}V_{2}=0$ is satisfied again and therefore
only the differential and the two-product are non-vanishing. They
are given by 
\begin{equation}
\tilde{p}_{W}V{}_{1}\tilde{i}_{W}=Q,\,\,\,\,\,\,\,\,\,\,\,\tilde{p}_{W}V{}_{2}\tilde{i}_{W}^{\otimes2}=m_{2}
\end{equation}
hence we indeed end up with Witten theory if we take the bilinear
form on the target space to be $\omega$. 

For the generalization to the sliver frame stubs we propose the following
maps:

\begin{equation}
\tilde{i}'_{W}=\begin{pmatrix}e^{-\lambda\mathcal{L}_{0}^{*}}\\
e^{-\lambda\mathcal{L}_{0}}e^{-\lambda\mathcal{L}_{0}^{*}}-1
\end{pmatrix},\,\,\,\,\,\,\,\,\,\,\tilde{p}'_{W}=\begin{pmatrix}e^{-\lambda\mathcal{L}_{0}} & -1\end{pmatrix}.\label{eq:cool i and p-1}
\end{equation}
\begin{equation}
\tilde{h}'_{W}=\frac{\mathcal{B}_{0}+\mathcal{B}_{0}^{*}}{\mathcal{L}_{0}+\mathcal{L}_{0}^{*}}\begin{pmatrix}e^{-\lambda\mathcal{L}_{0}^{*}}e^{-\lambda\mathcal{L}_{0}}-1 & -e^{-\lambda\mathcal{L}_{0}^{*}}\\
e^{-\lambda\mathcal{L}_{0}}e^{-\lambda\mathcal{L}_{0}^{*}}e^{-\lambda\mathcal{L}_{0}}-e^{-\lambda\mathcal{L}_{0}} & -e^{-\lambda\mathcal{L}_{0}}e^{-\lambda\mathcal{L}_{0}^{*}}
\end{pmatrix}.
\end{equation}
Again, $\tilde{i}'_{W}$, $\tilde{p}'_{W}$ and $\tilde{h}'_{W}$
form an SDR and obey $\tilde{h}'{}_{W}V'_{2}=0$, so we can directly
transfer to Witten theory: 
\begin{equation}
\tilde{p}'{}_{W}V'{}_{1}\tilde{i}'{}_{W}=Q,\,\,\,\,\,\,\,\,\,\,\,\tilde{p}'{}_{W}V'{}_{2}\tilde{i}'{}_{W}^{\otimes2}=m_{2}.
\end{equation}

\subsection{Cohomomorphisms and classical solutions}

We want to study now how to obtain classical solutions of the curly
auxiliary field theory, i. e. solve the equations of motion derived
from (\ref{eq:curly auxiliary action}). In coalgebra notation they
read 
\begin{equation}
\left(\mathbf{V'_{1}+V'_{2}}\right)\frac{1}{1-\Phi'}\equiv\mathbf{V'}\frac{1}{1-\Phi'}=0,
\end{equation}
where the schematic notation 
\begin{equation}
\frac{1}{1-\Phi'}=:\sum_{n=0}^{\infty}\Phi'{}^{\otimes n}
\end{equation}
is used. Given an analytic solution $\Psi$ of Witten theory, for
example the tachyon vacuum, we can ask if there exists a non-linear
map $A$ such that $A\left(\Psi\right)$ solves the equation above.
A sufficient condition is that $\mathbf{A}$ is a cohomomorphism that
intertwines between the algebras of Witten theory and the curly auxiliary
theory, i. e. 
\begin{equation}
\mathbf{V'A=Am}.
\end{equation}
for $\mathbf{m\equiv Q+m_{2}}.$ This relation arises as the chain
map relation automatically obeyed by the interacting inclusion $\mathbf{\tilde{I}'_{W}}$
coming from the homological perturbation lemma. Using the explicit
expression for $\mathbf{\tilde{I}'_{W}}$ we can directly observe
that 
\begin{equation}
\mathbf{\tilde{I}'{}_{W}}=\left(1-\mathbf{\tilde{h}'{}_{W}V'{}_{2}}\right)^{-1}\mathbf{\tilde{\mathbf{i}}'{}_{W}}=\mathbf{\tilde{\mathbf{i}}'{}_{W}}
\end{equation}
because of the magical relation $\mathbf{\tilde{h}'{}_{W}V'{}_{2}}=0$.
So using (\ref{eq:cool i and p-1}) we conclude that 
\begin{equation}
\Phi=\begin{pmatrix}e^{-\lambda\mathcal{L}_{0}^{*}}\Psi\\
e^{-\lambda\mathcal{L}_{0}}e^{-\lambda\mathcal{L}_{0}^{*}}\Psi-\Psi
\end{pmatrix}\label{eq:explicit two components}
\end{equation}
is a solution of the curly auxiliary field theory if $\Psi$ is a
solution of Witten theory. By direct computation we can see that $\mathbf{\tilde{I}'{}_{W}}$
is also cyclic, i. e. it preserves the action:\footnote{Cyclicity of a cohomomorphism is actually a delicate question and
was discussed in more detail in \cite{Schnabl2024}.} 
\begin{equation}
S=-\frac{1}{2}\Omega\left(\Phi,V'_{1}\left(\Phi\right)\right)-\frac{1}{3}\Omega\left(\Phi,V'_{2}\left(\Phi,\Phi\right)\right)=-\frac{1}{2}\omega\left(\Psi,Q\left(\Psi\right)\right)-\frac{1}{3}\omega\left(\Psi,m_{2}\left(\Psi,\Psi\right)\right).\label{eq:action equation}
\end{equation}

We can also ask if it is possible to transfer directly from Witten
theory to the curly-stubbed theory as it is discussed in \cite{Schnabl2024}.
For this we must combine the two homotopy transfers, so first include
from Witten theory to the curly auxiliary field theory and then project
down to the stubbed theory. The combined map is hence the product
$\mathbf{P'_{S}\tilde{I}'_{W}}$, where $\mathbf{P'_{S}}=\mathbf{p'_{S}}\left(1-\mathbf{V'_{2}h'_{S}}\right)^{-1}$
is the interacting projection of the homotopy transfer to the stubbed
theory. We can compute explicitly
\begin{align}
\pi_{1}\mathbf{P'_{S}\tilde{I}'{}_{W}}\pi_{1} & =p'_{S}\tilde{i}'{}_{W}=e^{-\lambda\mathcal{L}_{0}^{*}}\nonumber \\
\pi_{1}\mathbf{P'_{S}\tilde{I}'{}_{W}}\pi_{2}=p'{}_{S}V'{}_{2}h'{}_{S}\tilde{i}'{}_{W}^{\otimes2}= & e^{-\lambda\mathcal{L}_{0}^{*}}m_{2}\left(\cdot,\frac{\mathcal{B}_{0}+\mathcal{B}_{0}^{*}}{\mathcal{L}_{0}+\mathcal{L}_{0}^{*}}\left(e^{-\lambda\mathcal{L}_{0}}e^{-\lambda\mathcal{L}_{0}^{*}}-1\right)\cdot\right)\nonumber \\
 & +e^{-\lambda\mathcal{L}_{0}^{*}}m_{2}\left(\frac{\mathcal{B}_{0}+\mathcal{B}_{0}^{*}}{\mathcal{L}_{0}+\mathcal{L}_{0}^{*}}\left(e^{-\lambda\mathcal{L}_{0}}e^{-\lambda\mathcal{L}_{0}^{*}}-1\right)\cdot,e^{-\lambda\mathcal{L}_{0}}e^{-\lambda\mathcal{L}_{0}^{*}}\cdot\right)
\end{align}
and observe that $\mathbf{P'_{S}\tilde{I}'{}_{W}}$ is exactly equal
to the non-cyclic cohomomorphism $\mathbf{P}$ constructed in \cite{Schnabl2024}
up to second order. In the appendix we give a full proof that they
are equal to all orders. 

\section{Curly quantum stubs}

In this section we want to generalize the curly stubbed theory to
the full quantum level as it was done in \cite{Maccaferri2024} for
BPZ-even stubs. For that we need to turn the $A_{\infty}$-algebra
into a quantum $A_{\infty}$-algebra, i. e. the action has to obey
the quantum BV master equation
\begin{equation}
\frac{1}{2}\left(S,S\right)_{BV}+\Delta S=0
\end{equation}
where $\left(\cdot,\cdot\right)_{BV}$ is the BV-bracket and $\Delta$
is the BV-Laplacian. 

Suppose we are given any action of the form 
\begin{equation}
S\left(\Psi\right)=\sum_{n=1}^{\infty}\frac{1}{n+1}\omega\left(\Psi,m_{n}\left(\Psi^{\otimes n}\right)\right)
\end{equation}
with some higher products $m_{n}$ and a bilinear form $\omega.$
If we package the products into a coderivation as $\mathbf{m}=\mathbf{m_{1}}+\mathbf{m_{2}}+...$,
then $S$ can be written in the form 
\begin{equation}
S\left(\Psi\right)=\int_{0}^{1}dt\,\omega\left(\pi_{1}\boldsymbol{\boldsymbol{\partial_{t}}}\frac{1}{1-\Psi\left(t\right)},\pi_{1}\mathbf{m}\frac{1}{1-\Psi\left(t\right)}\right)
\end{equation}
where $\Psi\left(t\right)$ is a smooth interpolation between $\Psi\left(0\right)=0$
and $\Psi\left(1\right)=\Psi$, see for instance \cite{Vosmera:2020jyw}.
Using this representation, the BV master equation can be shown to
take the following form (see \cite{Maccaferri2023}):
\begin{equation}
\int_{0}^{1}dt\,\omega\left(\pi_{1}\boldsymbol{\boldsymbol{\partial_{t}}}\frac{1}{1-\Psi\left(t\right)},\pi_{1}\left(\mathbf{m}^{2}+\mathbf{mU}\right)\frac{1}{1-\Psi\left(t\right)}\right)=0.
\end{equation}
Here, $\mathbf{U}$ is the tensor coalgebra extension of the Poisson
bi-vector 
\begin{equation}
U=\frac{\left(-\right)^{v^{a}}}{2}v_{a}\wedge v^{a},
\end{equation}
where the $v_{a}$ provide a basis of $\mathcal{H}.$ The basis vectors
are taken to be canonically normalized via $\omega\left(v^{a},v_{b}\right)=-\omega\left(v_{b},v^{a}\right)=\delta_{b}^{a}$,
which means that $U$ implicitly depends on $\omega$. $\mathbf{U}$
is then defined by inserting $v_{a}$ and $v$$^{a}$ in all possible
combinations of positions in a tensor product. This can be understood
from the fact that $U$ is the dual object to the BV-Laplacian $\Delta$:
Since $\Delta$ is a second order differential operator, it removes
two elements from a tensor product. Similarly, $U$ inserts two elements
into a tensor product. $\Delta$ is not a derivation with respect
to the tensor product and likewise $\mathbf{U}$ is not a coderivation
on the tensor coalgebra and does not obey the co-Leibniz rule. 

If we first consider ordinary Witten theory, we immediately notice
that $\mathbf{m}^{2}=0$ already. The only non-trivial contribution
of the second term is $m_{2}\left(U\right)$, which is however actually
ill-defined and non-zero \cite{Thorn1989}. This resembles the fact
that pure open string field theory is incomplete on the quantum level
because it does not include the emission of closed strings. The inconsistency
is removed in the full quantized open-closed string field theory.
Since we are purely interested in the open sector, we will just formally
demand 
\begin{equation}
m_{2}\left(U\right)=:0\label{eq:quantum consistency condition}
\end{equation}
as a quantum consistency condition. 

Now let us consider the extended action (\ref{eq:curly auxiliary action}):
Again we know that $\mathbf{V'}^{2}=\left(\mathbf{V'_{1}}+\mathbf{V'_{2}}\right)^{2}=0$
so we only have to examine $V'_{2}\left(U'\right)$. First, we choose
a basis of $\mathcal{H}\oplus\mathcal{H}$ as
\begin{equation}
\left\{ \begin{pmatrix}v_{a}\\
0
\end{pmatrix},\begin{pmatrix}0\\
w_{b}
\end{pmatrix}\right\} ,
\end{equation}
where $v_{a}$ and $w_{b}$ can in principle be independent. By looking
at the symplectic form $\Omega'$ (\ref{eq:new symplectic form})
we see that the basis vectors of the auxiliary field change non-trivially
under dualization. In the end we get 
\begin{equation}
U'=\frac{\left(-\right)^{v^{a}}}{2}v_{a}\wedge v^{a}+\frac{\left(-\right)^{w^{a}}}{2}w_{a}\wedge\left(1-e^{-\lambda\mathcal{L}_{0}}e^{-\lambda\mathcal{L}_{0}^{*}}\right)w^{a}
\end{equation}
which can be directly inserted into 
\begin{equation}
V'_{2}\left(U'\right)=\begin{pmatrix}e^{-\lambda\mathcal{L}_{0}^{*}}\left(\frac{\left(-\right)^{v^{a}}}{2}m_{2}\left(e^{-\lambda\mathcal{L}_{0}}v_{a}\wedge e^{-\lambda\mathcal{L}_{0}}v^{a}\right)+\frac{\left(-\right)^{w^{a}}}{2}m_{2}\left(w_{a}\wedge\left(1-e^{-\lambda\mathcal{L}_{0}}e^{-\lambda\mathcal{L}_{0}^{*}}\right)w^{a}\right)\right)\\
\left(e^{-\lambda\mathcal{L}_{0}}e^{-\lambda\mathcal{L}_{0}^{*}}-1\right)\left(\frac{\left(-\right)^{v^{a}}}{2}m_{2}\left(e^{-\lambda\mathcal{L}_{0}}v_{a}\wedge e^{-\lambda\mathcal{L}_{0}}v^{a}\right)+\frac{\left(-\right)^{w^{a}}}{2}m_{2}\left(w_{a}\wedge\left(1-e^{-\lambda\mathcal{L}_{0}}e^{-\lambda\mathcal{L}_{0}^{*}}\right)w^{a}\right)\right)
\end{pmatrix}.
\end{equation}

We can relate the two bases of $\mathcal{H}$ according to 
\begin{equation}
w_{a}=e^{-\lambda\mathcal{L}_{0}}v_{a}
\end{equation}
This is a well-defined basis change; if we take for instance Fock
states as basis states, the operator $e^{-\lambda\mathcal{L}_{0}}$
just produces a number. From the normalization condition $\omega\left(v^{a},v_{b}\right)=\delta_{b}^{a}$
it follows that 
\begin{equation}
v^{a}=e^{-\lambda\mathcal{L}_{0}^{*}}w^{a}
\end{equation}
and we see that the term in parentheses becomes 
\begin{equation}
\frac{\left(-\right)^{w^{a}}}{2}m_{2}\left(w_{a}\wedge e^{-\lambda\mathcal{L}_{0}}e^{-\lambda\mathcal{L}_{0}^{*}}w^{a}\right)+\frac{\left(-\right)^{w^{a}}}{2}m_{2}\left(w_{a}\wedge\left(1-e^{-\lambda\mathcal{L}_{0}}e^{-\lambda\mathcal{L}_{0}^{*}}\right)w^{a}\right)=\frac{\left(-\right)^{w^{a}}}{2}m_{2}\left(w_{a}\wedge w^{a}\right).
\end{equation}
This is precisely the original quantum consistency condition, so we
have shown that quantizing the theory with sliver frame stubs does
not require any additional constraints.

Integrating out the auxiliary field works very similarly to the classical
case: We start with the SDR between $\mathcal{H}\oplus\mathcal{H}$
and $\mathcal{H}$ given by the natural inclusion and projection maps
(\ref{eq:natural i p}) and the homotopy (\ref{eq:h in B-hat}) and
lift it to the tensor coalgebra as usual. The difference is now that
we perturb not only by the product $\mathbf{V'_{2}}$, but by the
combination $\mathbf{V'_{2}}+\mathbf{U'}$. By the homological perturbation
lemma we get for the quantum higher products 
\begin{equation}
\left(M'_{qu}\right)_{n}=\pi_{1}\mathbf{M'_{qu}}\pi_{n}=\pi_{1}\left(\mathbf{Q}+\mathbf{p_{S}V'{}_{2}}\frac{1}{1+\mathbf{h'_{S}V'{}_{2}+h'{}_{S}U'}}\mathbf{i_{S}}\right)\pi_{n}.
\end{equation}
Note that although $\mathbf{U'}$ is not a coderivation, $\mathbf{M'_{qu}}$
is, since it obeys the co-Leibniz rule\footnote{GS wants to thank Jojiro Totsuka-Yoshinaka for discussions on that,
see also \cite{Konosu2025}}. However, in general it does not square to zero because only 
\begin{equation}
\left(\mathbf{M'_{qu}}+\mathbf{p_{S}U'}\right)^{2}=0
\end{equation}
has to be true, see \cite{Maccaferri2024}. As a result, the quantum
products $\left(M'_{qu}\right)_{n}$ do not form an ordinary $A_{\infty}$-algebra
but a quantum $A_{\infty}$-algebra, which contains extra higher loop
terms. The geometrical interpretation of that is that after introducing
stubs, the moduli space of higher genus Riemann surfaces with boundary
is not covered anymore by the Feynman diagrams. One needs extra vertices
with intrinsic loops in the action, i.e. loops that do not arise from
the Feynman rules. 

\section{Explicit solutions}

One of the main motivations of introducing stubs in OSFT is to study
the structure of solutions of $A_{\infty}$-theories, so in this section
we want to apply our results to some known classical solutions. 

\subsection{Tachyon vacuum}

Let us first consider the tachyon vacuum given by 
\begin{equation}
\Psi_{TV}=e^{-\frac{K}{2}}c\frac{KB}{1-e^{-K}}ce^{-\frac{K}{2}}
\end{equation}
with the elements of the $KBc$-algebra defined for instance in \cite{Erler:2019vhl,Okawa2006}.\footnote{Within calculations using $KBc$-elements we will write the Witten
product $m_{2}$ just as a normal product, according to the literature.} It obeys the sliver gauge condition 
\begin{equation}
\mathcal{B}_{0}\Psi_{TV}=0.
\end{equation}
In order to calculate $\Phi'$ we first need to know the action of
$e^{-\lambda\mathcal{L}_{0}^{*}}$ on $\Psi_{TV}.$ The calculation
was done in detail in \cite{Schnabl2024}; what basically happens
is that the width of the security strips $e^{-\frac{K}{2}}$ changes
and the $K$ in the denominator gets rescaled by $e^{\lambda}$. The
result is 
\begin{equation}
e^{-\lambda\mathcal{L}_{0}^{*}}\Psi_{TV}\,=e^{-K\left(e^{\lambda}-\frac{1}{2}\right)}c\frac{KB}{1-e^{-e^{\lambda}K}}ce^{-K\left(e^{\lambda}-\frac{1}{2}\right)}.
\end{equation}
For the second component we use again the formulas (\ref{L_hat exponential 1})
and (\ref{eq:L_hat exponential 2}) to get 
\begin{equation}
e^{-\lambda\mathcal{L}_{0}}e^{-\lambda\mathcal{L}_{0}^{*}}\Psi_{TV}=e^{-K\left(\frac{3}{2}-e^{-\lambda}\right)}c\frac{KB}{1-e^{-K}}ce^{-K\left(\frac{3}{2}-e^{-\lambda}\right)}.
\end{equation}
Now we can write down the full result as 
\begin{equation}
\Phi{}_{TV}=\begin{pmatrix}e^{-K\left(e^{\lambda}-\frac{1}{2}\right)}c\frac{KB}{1-e^{-e^{\lambda}K}}ce^{-K\left(e^{\lambda}-\frac{1}{2}\right)}\\
e^{-K\left(\frac{3}{2}-e^{-\lambda}\right)}c\frac{KB}{1-e^{-K}}ce^{-K\left(\frac{3}{2}-e^{-\lambda}\right)}-e^{-\frac{K}{2}}c\frac{KB}{1-e^{-K}}ce^{-\frac{K}{2}}
\end{pmatrix}.
\end{equation}
This result is remarkably simple compared to the expression for the
tachyon vacuum in the non-polynomial stubbed theory, cf. formula 4.24
in \cite{Schnabl2024}. The complexity comes in only after integrating
out the auxiliary field. 

\subsection{Identity-like solutions}

Another interesting application of stubs is the possibility of taming
certain singularities appearing in Witten theory. Here we want to
take a short look at one example, namely the famous identity-like
solution 
\begin{equation}
\Psi_{IL}=c\left(1-K\right)
\end{equation}

It obeys the Witten equations of motion as can be checked easily:
\begin{equation}
Q\Psi_{IL}+m_{2}\left(\Psi_{IL},\Psi_{IL}\right)=cKc\left(1-K\right)+c\left(1-K\right)c\left(1-K\right)=cKc\left(1-K\right)-cKc\left(1-K\right)=0.
\end{equation}

In \cite{Arroyo2010} it was shown that $\Psi_{IL}$ is actually gauge
related to the tachyon vacuum in the form 
\begin{equation}
\Psi_{STV}=c\left(1+K\right)Bc\frac{1}{1+K}
\end{equation}
called simple tachyon vacuum, which was derived in \cite{Erler:2009uj}.
Indeed, using the gauge parameter 
\begin{equation}
X=1+cBK,\,\,\,\,\,\,\,\,\,\,\,\,\,\,X^{-1}=1-cBK\frac{1}{1+K}
\end{equation}
one can prove by direct computation that 
\begin{equation}
\Psi_{STV}=X\Psi_{IL}X^{-1}+XQX^{-1}.
\end{equation}
The problems start when one tries to compute the action: Since $\Psi_{IL}$
is represented as a worldsheet strip of width zero, one would have
to calculate a correlator on a cylinder of vanishing circumference.
The result then always depends on the chosen regularization and is
hence ambiguous. 

In a theory with stubs however, this problem is avoided so we have
good reasons to believe that all quantities are well-defined. In \cite{Erler:2019vhl}
a proposal for a criterion is made that ``good'' solutions should
obey: Every function of $K$ appearing in the string field should
be bounded for $K\geq0.$ This is clearly not the case for $\Psi_{IL}$.
If we calculate the associated $\Phi_{IL}$ according to the formula
(\ref{eq:explicit two components}) we find 
\begin{equation}
\Phi_{IL}=\begin{pmatrix}e^{-\lambda\mathcal{L}_{0}^{*}}c\left(1-K\right)\\
e^{-\lambda\mathcal{L}_{0}}e^{-\lambda\mathcal{L}_{0}^{*}}c\left(1-K\right)-c\left(1-K\right)
\end{pmatrix}=\begin{pmatrix}e^{-\frac{K}{2}(e^{\lambda}-1)}c\left(e^{-\lambda}-K\right)e^{-\frac{K}{2}(e^{\lambda}-1)}\\
e^{(e^{-\lambda}-1)K}c\left(1-K\right)e^{(e^{-\lambda}-1)K}-c\left(1-K\right)
\end{pmatrix}
\end{equation}
where we used again the formulas derived in \cite{Schnabl2024}. Although
some of the functions of $K$ are now exponentially damped, there
is still the $cK$-term in the $\Sigma$-field which is unbounded.
If we try to compute the action we see from (\ref{eq:action equation})
that we just end up with the Witten action and hence run into the
same problems as before. However, if we integrate out the auxiliary
field and transfer to the non-polynomial theory, the situation changes:
In section two we calculated the cohomomorphism we need to apply and
have shown that it is actually identical to $\mathbf{P}$ from \cite{Schnabl2024,Schnabl:2023dbv}.
To second order in the string field we get 
\begin{align}
\Psi'{}_{IL}\equiv\pi_{1}\mathbf{P}\frac{1}{1-\Psi_{IL}}=\, & e^{-\lambda\mathcal{L}_{0}^{*}}c\left(1-K\right)+e^{-\lambda\mathcal{L}_{0}^{*}}m_{2}\left(c\left(1-K\right),\frac{\mathcal{B}_{0}+\mathcal{B}_{0}^{*}}{\mathcal{L}_{0}+\mathcal{L}_{0}^{*}}\left(e^{-\lambda\mathcal{L}_{0}}e^{-\lambda\mathcal{L}_{0}^{*}}-1\right)c\left(1-K\right)\right)\nonumber \\
 & +e^{-\lambda\mathcal{L}_{0}^{*}}m_{2}\left(\frac{\mathcal{B}_{0}+\mathcal{B}_{0}^{*}}{\mathcal{L}_{0}+\mathcal{L}_{0}^{*}}\left(e^{-\lambda\mathcal{L}_{0}}e^{-\lambda\mathcal{L}_{0}^{*}}-1\right)c\left(1-K\right),e^{-\lambda\mathcal{L}_{0}}e^{-\lambda\mathcal{L}_{0}^{*}}c\left(1-K\right)\right).
\end{align}
To simplify the action of $\frac{1}{\mathcal{L}_{0}+\mathcal{L}_{0}^{*}}$
we use a Schwinger parameter and write 
\begin{align}
\frac{\left(e^{-\lambda\mathcal{L}_{0}}e^{-\lambda\mathcal{L}_{0}^{*}}-1\right)}{\mathcal{L}_{0}+\mathcal{L}_{0}^{*}}c\left(1-K\right) & =\int_{0}^{\infty}dt\,\left(e^{(e^{-\lambda}-1-t)\left(\mathcal{L}_{0}+\mathcal{L}_{0}^{*}\right)}c\left(1-K\right)-e^{-t\left(\mathcal{L}_{0}+\mathcal{L}_{0}^{*}\right)}c\left(1-K\right)\right)\nonumber \\
 & =-\int_{0}^{1-e^{-\lambda}}dt\,e^{-tK}c\left(1-K\right)e^{-tK}.
\end{align}
Together with the well-known formula 
\begin{equation}
\left(\mathcal{B}_{0}+\mathcal{B}_{0}^{*}\right)\Psi=B\Psi+\left(-\right)^{\text{gh}\left(\Psi\right)}\Psi B
\end{equation}
we can write the final result as 
\begin{align}
\Psi'{}_{IL}=\, & e^{-\frac{K}{2}(e^{\lambda}-1)}c\left(e^{-\lambda}-K\right)e^{-\frac{K}{2}(e^{\lambda}-1)}\nonumber \\
 & -e^{-\frac{K}{2}\left(e^{\lambda}-1\right)}c\left(e^{-\lambda}-K\right)\int_{0}^{1-e^{-\lambda}}dt\,e^{-e^{\lambda}tK}\left(Bc-cB\right)\left(1-e^{\lambda}K\right)e^{-\left(\left(t+\frac{1}{2}\right)e^{\lambda}-\frac{1}{2}\right)K}\nonumber \\
 & -\int_{0}^{1-e^{-\lambda}}dt\,e^{-\left(\left(t+\frac{1}{2}\right)e^{\lambda}-\frac{1}{2}\right)K}\left(Bc-cB\right)\left(1-e^{\lambda}K\right)e^{(1-e^{\lambda}-e^{\lambda}t)K}c\left(e^{-\lambda}-K\right)e^{-\left(\left(t+\frac{1}{2}\right)e^{\lambda}-e^{-\lambda}+\frac{1}{2}\right)K}\nonumber \\
 & +\mathcal{O}\left(\Psi^{3}\right).
\end{align}
We see that every function of $K$ appearing is now exponentially
damped and therefore bounded for positive $K$.\footnote{The only potential problem could come from the lower integration bound
in the second line. However, usually a continuous superposition of
states with width going to zero as it appears for instance in the
single tachyon vacuum \cite{Erler:2009uj} is much better behaved
than a single identity-like state. }Since the higher orders of the cohomomorphism are constructed out
of the same ingredients as the first two, we conclude that this result
is true to all orders. We could now plug in the result into the stubbed
action and get 
\begin{equation}
S=-\frac{1}{2}\omega\left(\Psi'_{IL},Q\Psi'_{IL}\right)-\frac{1}{3}\omega\left(\Psi'_{IL},M_{2}\left(\Psi'_{IL},\Psi'_{IL}\right)\right)-\frac{1}{4}\omega\left(\Psi'_{IL},M_{3}\left(\Psi'_{IL},\Psi'_{IL},\Psi'_{IL}\right)\right)-\cdots
\end{equation}
with the sliver frame products defined in \cite{Schnabl2024}. This
expression does not contain any singular correlators anymore and should
hence yield a well-defined result. In \cite{Schnabl2024} it has been
argued that the on-shell value of the action is preserved even for
the non-cyclic cohomomorphism $\mathbf{P}$, so we expect the result
to be the total energy of the tachyon vacuum, namely $-\frac{1}{2\pi^{2}}.$
The calculation can in principle be done up to some finite order to
check the claim and also study the convergence properties, but this
goes beyond the scope of this paper. 

\section{Conclusion}

In this paper we succeeded to extend the formalism developed by Erbin
and Firat to non-BPZ-even stubs, in particular the sliver frame stub.
We have also shown, following the route outlined by Maccaferri et
al, that the sliver frame stubbed theory is consistent at the full
quantum level without any additional assumptions. By applying our
construction to the tachyon vacuum, we could see that its auxiliary
field form is remarkably simple, in contrast to its expression in
the stubbed theory. Moreover, the identity-like solution $c\left(1-K\right)$
takes a non-singular form in the stubbed theory and gives rise to
an action which is computable unambiguously. For future work it would
be of interest to calculate this action up to some finite order and
study its value and convergence properties. In general, examining
the structure of solutions via the auxiliary field method can teach
us important algebraic properties which can be relevant also in different
context. For example, it might be possible to formulate other theories
like open superstring field theory \cite{Erler2014} or even closed
string field theory \cite{Zwiebach1993} in a cubic way using auxiliary
fields. 

\subsubsection*{\textcolor{black}{Acknowledgements}}

\textcolor{black}{GS wants to thank Martin Schnabl for suggesting
the topic and collaboration as well as Carlo Maccaferri, Alberto Ruffino,
Harold Erbin, Atakan Firat and }Jojiro Totsuka-Yoshinaka\textcolor{black}{{}
for useful discussions. Our work has been funded by the Grant Agency
of Czech Republic under the grant EXPRO 20-25775X.}

\appendix

\section{Appendix }

In this appendix we want to prove the statement of the end of section
2.3, that $\mathbf{P'_{S}\tilde{I}'_{W}}$ is equal to the cohomomorphism
$\mathbf{P}$ from \cite{Schnabl2024} to all orders. The proof closely
follows a related one in the appendix of \cite{Erbin2023}. 

The first step is to write the product $V'_{2}$ as 
\begin{equation}
V'_{2}=\tilde{i}'{}_{W}m_{2}\tilde{p}'{}_{W}^{\otimes2}
\end{equation}
using the maps $\tilde{i}'_{W}$ and $\tilde{p}'_{W}$ we used earlier
to project directly onto Witten theory. The equation can be lifted
to the tensor coalgebra in the following way: 
\begin{equation}
\mathbf{V'_{2}=\Pi\left(\tilde{i}'_{W}m_{2}\tilde{p}'_{W}\right)}.
\end{equation}
Here, $\Pi$ is a formal object that sets the combination $\tilde{i}'_{W}\tilde{p}'_{W}$
equal to unity whenever it appears in any tensor power. We can write
the cohomomorphism $\mathbf{P'_{S}\tilde{I}'_{W}}$ now as 
\begin{equation}
\mathbf{P'_{S}\tilde{I}'_{W}}=\mathbf{p'_{S}}\left(1-\mathbf{\Pi\left(\tilde{i}'_{W}m_{2}\tilde{p}'_{W}\right)h'_{S}}\right)^{-1}\mathbf{\tilde{i}'_{W}}\label{eq:eq with Pi}
\end{equation}
and observe some relations: First, the combination $\tilde{p}'_{W}h'_{S}\tilde{i}'_{W}$
is equal to 
\begin{equation}
\tilde{p}'_{W}h'_{S}\tilde{i}'_{W}=\begin{pmatrix}e^{-\lambda\mathcal{L}_{0}} & -1\end{pmatrix}\begin{pmatrix}0 & 0\\
0 & -\frac{\mathcal{B}_{0}+\mathcal{B}_{0}^{*}}{\mathcal{L}_{0}+\mathcal{L}_{0}^{*}}
\end{pmatrix}\begin{pmatrix}e^{-\lambda\mathcal{L}_{0}^{*}}\\
e^{-\lambda\mathcal{L}_{0}}e^{-\lambda\mathcal{L}_{0}^{*}}-1
\end{pmatrix}=\frac{e^{-\lambda\mathcal{L}_{0}}e^{-\lambda\mathcal{L}_{0}^{*}}-1}{\mathcal{L}_{0}+\mathcal{L}_{0}^{*}}\left(\mathcal{B}_{0}+\mathcal{B}_{0}^{*}\right).
\end{equation}
Second, the combination $i_{S}p_{S}$ that also appears in the tensor
algebra quantity $\mathbf{h'_{S}}$ combines with $\mathbf{\tilde{i}'_{W}}$
and $\mathbf{\tilde{p}'_{W}}$ to 
\begin{equation}
\tilde{p}'_{W}i_{S}p_{S}\tilde{i}'_{W}=\begin{pmatrix}e^{-\lambda\mathcal{L}_{0}} & -1\end{pmatrix}\begin{pmatrix}1 & 0\\
0 & 0
\end{pmatrix}\begin{pmatrix}e^{-\lambda\mathcal{L}_{0}^{*}}\\
e^{-\lambda\mathcal{L}_{0}}e^{-\lambda\mathcal{L}_{0}^{*}}-1
\end{pmatrix}=e^{-\lambda\mathcal{L}_{0}}e^{-\lambda\mathcal{L}_{0}^{*}}.
\end{equation}
This implies that we can write 
\begin{equation}
\mathbf{\tilde{p}'_{W}h'_{S}\mathbf{\tilde{i}'_{W}}}=\underset{n=1}{\overset{\infty}{\sum}}\underset{k=0}{\overset{n-1}{\sum}}1^{\otimes k}\otimes\frac{e^{-\lambda\mathcal{L}_{0}}e^{-\lambda\mathcal{L}_{0}^{*}}-1}{\mathcal{L}_{0}+\mathcal{L}_{0}^{*}}\left(\mathcal{B}_{0}+\mathcal{B}_{0}^{*}\right)\otimes\left(e^{-\lambda\mathcal{L}_{0}}e^{-\lambda\mathcal{L}_{0}^{*}}\right)^{\otimes n-k-1}.
\end{equation}
which is precisely equal to the homotopy in $\hat{\mathcal{B}}$-gauge
homotopy $\mathbf{h_{\boldsymbol{\hat{\mathcal{B}}}}}$ defined in
\cite{Schnabl2024}. Moreover, $p'_{S}\tilde{i}'_{W}=e^{-\lambda\mathcal{L}_{0}^{*}}$
and hence equal to the projection $p$ used in \cite{Schnabl2024}.
We see that if we could ignore the projector $\Pi$ for a moment,
we could write 
\begin{equation}
\mathbf{P'_{S}\tilde{I}'_{W}}\overset{?}{=}\mathbf{p}\left(1-\mathbf{m_{2}}\mathbf{h_{\boldsymbol{\hat{\mathcal{B}}}}}\right)^{-1}
\end{equation}
which can be compared to 
\begin{equation}
\mathbf{P}=P_{SDR}\left(\mathbf{p\left(1-m_{2}\mathbf{h_{\boldsymbol{\hat{\mathcal{B}}}}}\right)^{-1}}\right).
\end{equation}

Here, $P_{SDR}$ is a formal object that implements the SDR-conditions
$pi=1$, $h_{\mathcal{\hat{B}}}i=ph_{\mathcal{\hat{B}}}=h_{\mathcal{\hat{B}}}h_{\mathcal{\hat{B}}}=0$
($h_{\mathcal{\hat{B}}}h_{\mathcal{\hat{B}}}=0$ is already obeyed),
for a more precise definition see \cite{Schnabl:2023dbv}. We see
that the expressions are equal except for the projectors $\Pi$ and
$P_{SDR}$, so we have to find out how to relate them. Let us look
at the individual terms explicitly and observe the effect of $\Pi$:
As usual it is enough to consider the projection on one output. For
one or two inputs we do not get any constraints, so the first non-trivial
term without $\Pi$ would be 
\begin{align}
\pi_{1}\mathbf{P'_{S}\tilde{I}'_{W}}\pi_{3}\overset{?}{=}\, & p'_{S}\tilde{i}'_{W}m_{2}\left(\tilde{p}'_{W}\otimes\tilde{p}'_{W}\right)\left(\mathds{1}\otimes h'_{S}+h'_{S}\otimes i_{S}p_{S}\right)\left(\tilde{i}'_{W}\otimes\tilde{i}'_{W}\right)\left(\mathds{1}\otimes m_{2}+m_{2}\otimes\mathds{1}\right)\nonumber \\
 & \left(\tilde{p}'_{W}\otimes\tilde{p}'_{W}\otimes\tilde{p}'_{W}\right)\left(\mathds{1}\otimes\mathds{1}\otimes h'_{S}+\mathds{1}\otimes h'_{S}\otimes i_{S}p_{S}+h'_{S}\otimes i_{S}p_{S}\otimes i_{S}p_{S}\right)\left(\tilde{i}'_{W}\otimes\tilde{i}'_{W}\otimes\tilde{i}'_{W}\right)
\end{align}
The terms containing $\tilde{i}'_{W}\tilde{p}'_{W}$ are 
\begin{align}
 & p'_{S}\tilde{i}'_{W}m_{2}\left(\tilde{p}'_{W}\otimes\tilde{p}'_{W}\right)\left(\mathds{1}\otimes h'_{S}+h'_{S}\otimes i_{S}p_{S}\right)\left(\tilde{i}'_{W}\tilde{p}'_{W}\otimes\tilde{i}'_{W}m_{2}\left(\tilde{p}'_{W}\otimes\tilde{p}'_{W}\right)+\tilde{i}'_{W}m_{2}\left(\tilde{p}'_{W}\otimes\tilde{p}'_{W}\right)\otimes\tilde{i}'_{W}\tilde{p}'_{W}\right)\nonumber \\
 & \left(\mathds{1}\otimes\mathds{1}\otimes h'_{S}+\mathds{1}\otimes h'_{S}\otimes i_{S}p_{S}+h'_{S}\otimes i_{S}p_{S}\otimes i_{S}p_{S}\right)\left(\tilde{i}'_{W}\otimes\tilde{i}'_{W}\otimes\tilde{i}'_{W}\right)
\end{align}
and we see that due to $\tilde{p}'_{W}\tilde{i}'_{W}=1$ in many of
the terms it does not make any difference if $\tilde{i}'_{W}\tilde{p}'_{W}$
is replaced by unity. The problematic terms are 
\[
p'_{S}\tilde{i}'_{W}m_{2}\left(\tilde{p}'_{W}\otimes\tilde{p}'_{W}\right)\left(h'_{S}\otimes i_{S}p_{S}\right)\left(\tilde{i}'_{W}\tilde{p}'_{W}\otimes\tilde{i}'_{W}m_{2}\left(\tilde{p}'_{W}\otimes\tilde{p}'_{W}\right)\right)\left(h'_{S}\otimes i_{S}p_{S}\otimes i_{S}p_{S}\right)\left(\tilde{i}'_{W}\otimes\tilde{i}'_{W}\otimes\tilde{i}'_{W}\right),
\]
\[
p'_{S}\tilde{i}'_{W}m_{2}\left(\tilde{p}'_{W}\otimes\tilde{p}'_{W}\right)\left(h'_{S}\otimes i_{S}p_{S}\right)\left(\tilde{i}'_{W}m_{2}\left(\tilde{p}'_{W}\otimes\tilde{p}'_{W}\right)\otimes\tilde{i}'_{W}\tilde{p}'_{W}\right)\left(\mathds{1}\otimes\mathds{1}\otimes h'_{S}\right)\left(\tilde{i}'_{W}\otimes\tilde{i}'_{W}\otimes\tilde{i}'_{W}\right),
\]
\[
p'_{S}\tilde{i}'_{W}m_{2}\left(\tilde{p}'_{W}\otimes\tilde{p}'_{W}\right)\left(h'_{S}\otimes i_{S}p_{S}\right)\left(\tilde{i}'_{W}m_{2}\left(\tilde{p}'_{W}\otimes\tilde{p}'_{W}\right)\otimes\tilde{i}'_{W}\tilde{p}'_{W}\right)\left(\mathds{1}\otimes h'_{S}\otimes i_{S}p_{S}\right)\left(\tilde{i}'_{W}\otimes\tilde{i}'_{W}\otimes\tilde{i}'_{W}\right),
\]
\[
p'_{S}\tilde{i}'_{W}m_{2}\left(\tilde{p}'_{W}\otimes\tilde{p}'_{W}\right)\left(h'_{S}\otimes i_{S}p_{S}\right)\left(\tilde{i}'_{W}m_{2}\left(\tilde{p}'_{W}\otimes\tilde{p}'_{W}\right)\otimes\tilde{i}'_{W}\tilde{p}'_{W}\right)\left(h'_{S}\otimes i_{S}p_{S}\otimes i_{S}p_{S}\right)\left(\tilde{i}'_{W}\otimes\tilde{i}'_{W}\otimes\tilde{i}'_{W}\right),
\]
\[
p'_{S}\tilde{i}'_{W}m_{2}\left(\tilde{p}'_{W}\otimes\tilde{p}'_{W}\right)\left(\mathds{1}\otimes h'_{S}\right)\left(\tilde{i}'_{W}m_{2}\left(\tilde{p}'_{W}\otimes\tilde{p}'_{W}\right)\otimes\tilde{i}'_{W}\tilde{p}'_{W}\right)\left(\mathds{1}\otimes\mathds{1}\otimes h'_{S}\right)\left(\tilde{i}'_{W}\otimes\tilde{i}'_{W}\otimes\tilde{i}'_{W}\right),
\]
\[
p'_{S}\tilde{i}'_{W}m_{2}\left(\tilde{p}'_{W}\otimes\tilde{p}'_{W}\right)\left(\mathds{1}\otimes h'_{S}\right)\left(\tilde{i}'_{W}m_{2}\left(\tilde{p}'_{W}\otimes\tilde{p}'_{W}\right)\otimes\tilde{i}'_{W}\tilde{p}'_{W}\right)\left(\mathds{1}\otimes h'_{S}\otimes i_{S}p_{S}\right)\left(\tilde{i}'_{W}\otimes\tilde{i}'_{W}\otimes\tilde{i}'_{W}\right)
\]
and
\[
p'_{S}\tilde{i}'_{W}m_{2}\left(\tilde{p}'_{W}\otimes\tilde{p}'_{W}\right)\left(\mathds{1}\otimes h'_{S}\right)\left(\tilde{i}'_{W}m_{2}\left(\tilde{p}'_{W}\otimes\tilde{p}'_{W}\right)\otimes\tilde{i}'_{W}\tilde{p}'_{W}\right)\left(h'_{S}\otimes i_{S}p_{S}\otimes i_{S}p_{S}\right)\left(\tilde{i}'_{W}\otimes\tilde{i}'_{W}\otimes\tilde{i}'_{W}\right).
\]
The first and fifth term vanish because they contain $\left(\tilde{p}'_{W}h'_{S}\tilde{i}'_{W}\right)^{2}=h_{\mathcal{\hat{B}}}^{2}=0$.
The second term would actually vanish if $\tilde{i}'_{W}\tilde{p}'_{W}=\mathds{1}$
because of $p_{S}h'_{S}=0$. Otherwise we get the combination $\tilde{p}'_{W}i_{S}p_{S}\tilde{i}'_{W}\tilde{p}'_{W}h'_{S}\tilde{i}'_{W}$
which is equal to $iph_{\mathcal{\hat{B}}}$ in the notation of \cite{Schnabl2024}
and hence gets eliminated by $P_{SDR}$. In the third and fourth term
we would have $\tilde{p}'_{W}i_{S}p_{S}\tilde{i}'_{W}\tilde{p}'_{W}i_{S}p_{S}\tilde{i}'_{W}$
replaced by $\tilde{p}'_{W}i_{S}p_{S}\tilde{i}'_{W}$ which corresponds
to replacing $ipip$ by $ip$ as implemented by $P_{SDR}$. Similarly,
in the last two terms the combination $\tilde{p}'_{W}h'_{S}\tilde{i}'_{W}\tilde{p}'_{W}i_{S}p_{S}\tilde{i}'_{W}=h_{\mathcal{\hat{B}}}ip$
would vanish if $\tilde{i}'_{W}\tilde{p}'_{W}=\mathds{1}$ due to
$h'_{S}i_{S}=0$ whileas in the other formula $h_{\mathcal{\hat{B}}}i$
is deleted by $P_{SDR}$. The structure is the same for all orders
since no new combinations of operators appear so we can conclude that
the projectors $\Pi$ and $P_{SDR}$ have the same effect on all terms
that appear. This completes the proof of equivalence between $\mathbf{P'_{S}\tilde{I}'_{W}}$
and $\mathbf{P}$.

\bibliographystyle{plain}
\bibliography{String_Field_Theory}

\end{document}